\documentclass[a4paper, 11pt, oneside]{article}

\usepackage[english]{babel} 
\usepackage{microtype} 
\usepackage{xspace} 
\usepackage[utf8]{inputenc}	
\usepackage[T1]{fontenc} 

\usepackage{amsfonts,amsmath,amssymb,bm,mathtools,slashed} 

\usepackage{multicol} 
\usepackage[dvipsnames,table]{xcolor}	
\newcommand{\myColor}[0]{MidnightBlue}
\newcommand{\myColorLink}[0]{MidnightBlue}

\colorlet{darkBlue}{blue!45!black}
\colorlet{linkColor}{blue!80!black}
\usepackage{hyperref} 
\hypersetup{
	colorlinks, 
	bookmarksopen, 
	bookmarksnumbered,
	pdftoolbar=false, 
	pdfmenubar=false, 
	pdffitwindow=false, 
	pdfstartview={FitH},	 
	citecolor=\myColorLink, 
	linkcolor=\myColorLink,	
	urlcolor=\myColorLink, 
	pdftitle={Notes}, 
	pdfsubject={Notes}, 
	pdfauthor={Felix Wilsch}, 
	pdfnewwindow=true, 
	linktoc=page
}

\usepackage{geometry}
\usepackage{fancyhdr}

\fancypagestyle{main}{%
    \fancyhf{}
    
    \setcounter{page}{1}
    \pagenumbering{arabic}
	\cfoot{--~\thepage~--}
}

\fancypagestyle{intro}{%
    \fancyhf{}
    
    \setcounter{page}{1}
    \pagenumbering{roman}
	\cfoot{--~\thepage~--}
}

\usepackage{titlesec}
\titleformat{\section}{\color{\myColor}\Large\normalfont\bfseries}{\thesection}{1.0em}{}
\titleformat{\subsection}{\color{\myColor}\large\normalfont\bfseries}{\thesubsection}{1.0em}{}
\titleformat{\subsubsection}{\color{\myColor}\normalfont\bfseries}{\thesubsubsection}{1.0em}{}

\makeatletter
\g@addto@macro\bfseries{\boldmath} 
\makeatother

\usepackage[sort&compress,numbers,merge]{natbib}
\addto\captionsenglish{}

\usepackage{graphicx}
\usepackage{epstopdf}
\graphicspath{{./Figures/}{./figures/}}
\usepackage{caption,subcaption}
\usepackage{tikz}
\usepackage[compat=1.1.0]{tikz-feynman}
\usepackage{booktabs} 
\usepackage{multirow} 

\usepackage[shortlabels]{enumitem}
\setlist{itemsep=.1em,topsep=.5em}
\SetEnumerateShortLabel{i}{\textit{\roman*}}

\numberwithin{equation}{section} 
\allowdisplaybreaks 

\usepackage{float}

\usepackage{color, colortbl}
\definecolor{LightCyan}{rgb}{1,1,1}

\usepackage{soul}
\usepackage{comment}

\newcommand{\cmd}[1]{{\texttt{#1}}\xspace}
\newcommand{\dd}{\mathop{}\!\mathrm{d}}

\newcommand{\brackets}[1]{\left( #1 \right)}
\newcommand{\squarebrackets}[1]{\left[ #1 \right]} 

\newcommand{\abs}[1]{\left\lvert #1 \right\rvert}

\renewcommand{\L}{\mathcal{L}}
\newcommand{\ord}[1]{\mathcal{O}\!\left( #1 \right)}
\newcommand{\C}[2]{C_{\underset{#2}{#1}}}
\newcommand{\CL}[2]{\mathcal{C}_{\underset{#2}{#1}}}

\newcommand{\loopfactor}[0]{\frac{1}{16 \pi^2}\log\left(\frac{\mu_L}{\mu_H}\right)}
\newcommand{\be}{\begin{equation}}
\newcommand{\ee}{\end{equation}}
\newcommand{\bea}{\begin{eqnarray}}
\newcommand{\eea}{\end{eqnarray}}  
\newcommand{\no}{\nonumber}
\newcommand{\gsim}{\lower.7ex\hbox{$\;\stackrel{\textstyle>}{\sim}\;$}}
\newcommand{\lsim}{\lower.7ex\hbox{$\;\stackrel{\textstyle<}{\sim}\;$}}
\newcommand{\cO}{{\mathcal O}}
\newcommand{\cC}{{\mathcal C}}
\newcommand{\cL}{{\mathcal L}}

\begin{document}

\thispagestyle{empty}

\renewcommand*{\thefootnote}{\fnsymbol{footnote}} 

\begin{center} 
\begin{minipage}{15.5cm}
\vspace{-0.7cm}
\begin{flushright}
{\footnotesize \cmd{
ZU-TH-58/21
}}
\end{flushright}
\end{minipage}
\end{center}

\begin{center}
{
	\Large 
	\bfseries 
	Flavour alignment of New Physics \\ in light of the  $(g-2)_\mu$ anomaly
}
\\[0.8cm]
{ 
	Gino Isidori,\footnote{\href{mailto:isidori@physik.uzh.ch}{isidori@physik.uzh.ch}}
	Julie Pag\`es,\footnote{\href{mailto:julie.pages@physik.uzh.ch}{julie.pages@physik.uzh.ch}}
	and Felix Wilsch\footnote{\href{mailto:felix.wilsch@physik.uzh.ch}{felix.wilsch@physik.uzh.ch}}
}
\\[0.1cm]
{
	\small 
	Physik-Institut, Universit\"at Z\"urich, CH-8057 Z\"urich, Switzerland
}
\\[0.4cm]
{
	\it \today
}
\end{center}

\setcounter{footnote}{0}
\renewcommand*{\thefootnote}{\arabic{footnote}}%
\suppressfloats	

\begin{abstract}
We investigate the flavour alignment conditions that New Physics (NP) models need to satisfy in order to address the $(g-2)_\mu$ anomaly and, at the same time, be consistent with the tight bounds from $\mu \to e \gamma$ and $\tau \to \mu \gamma$.
We analyse the problem in general terms within the SMEFT, considering the renormalisation group
evolution of all the operators involved.
We show that semileptonic four-fermion operators, which are likely to generate a sizeable contribution to the $(g-2)_\mu$ anomaly, 
need to be tightly aligned to the lepton Yukawa couplings and the dipole operators in flavour space. 
While this tuning can be achieved in specific NP constructions, employing particular dynamical assumptions and/or flavour symmetry hypotheses, it is problematic in a wide class of models with broken flavour symmetries, such as those proposed to address both charged- and neutral-current $B$~anomalies. We quantify this tension both in general terms, and in the context of explicit NP constructions. 
\end{abstract}

\vskip 2 cm

\section{Introduction}
The anomalous magnetic moment of the muon, $a_\mu =(g_\mu-2)/2$, is a very powerful probe 
of possible physics beyond the Standard Model (SM). The recent experimental measurement of $a_\mu$
 by the E989 experiment at FNAL~\cite{Muong-2:2021ojo}, 
combined with the previous BNL result~\cite{Muong-2:2006rrc}, 
has strengthened the discrepancy with the SM prediction reported 
in Ref.~\cite{Aoyama:2020ynm} (see also \cite{Jegerlehner:2017gek,Jegerlehner:2017gek,Colangelo:2018mtw,Hoferichter:2019mqg,Davier:2019can,Keshavarzi:2019abf,Hoid:2020xjs,Czarnecki:2002nt, Melnikov:2003xd,Aoyama:2012wk, Gnendiger:2013pva}). While there is still some debate 
on the precise value of the SM prediction $a^{\rm SM}_\mu$  (see in particular Ref.~\cite{Borsanyi:2020mff}),
the recent experimental result has stimulated a renewed interest in possible 
beyond-the-SM (BSM) contributions to this observable. A~general analysis of $a_\mu$ 
within the SM Effective Field Theory (SMEFT), i.e.~under the hypothesis of new degrees of freedom
above the electroweak scale, has been presented in Ref.~\cite{Aebischer:2021uvt,Allwicher_2021,Allwicher:2021rtd,Fajfer:2021cxa}.

The interest in BSM contributions to $a_\mu$ is further reinforced 
by the deviations from lepton flavour universality observed in neutral-current~\cite{LHCb:2014vgu,LHCb:2017avl,LHCb:2019hip,LHCb:2021trn}
and charged-current~\cite{BaBar:2012obs,BaBar:2013mob,Belle:2015qfa,LHCb:2015gmp,LHCb:2017smo,LHCb:2017rln} semileptonic $B$~decays,
often referred to as the $B$~physics anomalies.
Indeed, also these observables indicate deviations from SM predictions in processes 
involving charged leptons and in particular muons (in the case of the neutral-current anomalies).
Extensions of the SM aiming at combined explanations of $a_\mu$ and (some of) the $B$~physics anomalies
have been presented in Ref.~\cite{Greljo:2021xmg,Baum:2021qzx,Lee:2021jdr,Arcadi:2021cwg,Cen:2021iwv,Marzocca:2021azj,Altmannshofer:2021hfu,Cacciapaglia:2021gff,Davighi:2021oel,Marzocca:2021miv,Greljo:2021npi,Bause:2021prv}, 
The $a_\mu$ anomaly involves only leptons of the second generation and, taken alone, does 
not indicate any violation of flavour quantum numbers. It also does not provide a clear indication 
on the energy scale of the associated new dynamics~\cite{Allwicher_2021,Allwicher:2021rtd}. 
On the other hand, the $B$~physics anomalies 
involve fermions of different generations (if combined),  necessarily implicate flavour changing dynamics 
(at least on the quark side), and point to New Physics above the electroweak scale. 
The goal of this paper is to analyse the general implications of the putative $a_\mu$ anomaly 
on the lepton flavour structure of the underlying NP model, assuming the latter is well described by the SMEFT.
A~key question to address in general terms is the compatibility of the $a_\mu$ anomaly with other 
non-standard phenomena and, moreover, to investigate the class of NP models favoured by this anomaly.

Within the SM, the three separate 
lepton flavour quantum numbers are conserved. However, this is an accidental property
of the $d=4$ operators of the SMEFT, which might well be violated in the ultraviolet (UV) 
completion of the theory. Indeed already at $d=5$, in the neutrino mass matrix, we observe 
large lepton flavour mixing effects. As we shall show using general EFT arguments, if the 
$a_\mu$ anomaly is confirmed as clear evidence of NP,  we are forced to assume that the 
conservation of lepton flavour  plays an important role also above the electroweak scale.
  More precisely, the interplay between the possible evidence for NP 
associated to the $a_\mu$ anomaly and the tight bounds from $\mu \to e \gamma$ 
and  $\tau \to \mu \gamma$, imply that lepton flavour (and in particular the  electron flavour) 
must be conserved to a very good accuracy in a whole sector of the $d=6$ SMEFT operators. 
As we shall discuss, this symmetry property of the SMEFT is unlikely to hold accidentally.

The paper is organised as follows: in Section~\ref{sect:exp} we briefly summarise the size of  
the muon dipole operator at the electroweak scale implied by the $a_\mu$ anomaly, and the 
bounds on the flavour violating dipole operators following from the non-observation of $\mu \to e \gamma$ 
and  $\tau \to \mu \gamma$. In Section~\ref{sect:RG} we show how, via Renormalisation Group~(RG)  
evolution, this information translates into constraints on other SMEFT operators at higher scales, 
implying a non-trivial series of alignment conditions in flavour space. 
In Section~\ref{sect:align} we discuss how to fulfil these alignment conditions in general terms,
either using dynamical hypotheses or via flavour symmetries.  An illustrative implementation
of these general mechanisms into a simple NP model based on scalar mediators is presented in 
Section~\ref{sect:mediators}. 
The results are summarised in the Conclusions.

\section{Experimental constraints on the leptonic dipole operators}
\label{sect:exp}

We work under the assumption  that  NP is heavy and can be well described by the  SMEFT Lagrangian,
including effective operators up to $d=6$. 
The focus of our analysis are the leptonic dipole operators 
\be
\mathcal{O}_{\underset{rs}{e\gamma}}
= \frac{v }{\sqrt{2}}  \overline{e}_{L_r}  \sigma^{\mu\nu} e_{R_s} F_{\mu\nu}
\label{eq:dipoledef}
\ee
written here below the scale of electroweak symmetry breaking,
where $r$ and~$s$ are generic flavour indices and $F_{\mu\nu}$ is the electromagnetic field strength tensor. Depending on the flavour structure, these operators 
can describe both NP effects in $a_\mu$, as well as non-vanishing rates for $\mu \to e \gamma$ and  $\tau \to \mu \gamma$.
The key point we want to investigate is the interplay between the evidence of a non-vanishing value for (some of) the Wilson 
coefficients of these operators, following from the $a_\mu$ anomaly, and the tight constraints derived 
by the non-observation of  $\mu \to e \gamma$ and  $\tau \to \mu \gamma$.

The combined result from the E989 experiment at FNAL~\cite{Muong-2:2021ojo} and the E821 experiment at BNL~\cite{Muong-2:2006rrc} on $a_\mu$, 
together with the SM prediction in~\cite{Aoyama:2020ynm}, imply 
\begin{align}
\Delta a_\mu &= a_\mu^\mathrm{Exp} - a_\mu^\mathrm{SM} = \brackets{251 \pm 59} \times 10^{-11}~.
\end{align}
The tree-level expression for  $\Delta a_\mu$ in terms of the Wilson coefficient of the dipole operator is 
\begin{align}
\Delta a_{\mu} &= \frac{4 m_{\mu}}{e}   \frac{v}{\sqrt 2}\text{Re} \,\CL{e\gamma}{22}^\prime \,,
\label{eq:magnetic-moment}
\end{align}
where $v=(\sqrt{2} G_F)^{-1/2} \approx 246$~GeV.
Here, the Wilson coefficient is understood to be evaluated at the weak scale 
(we neglect the small effect of running below the weak scale)
and the  prime  indicates the flavour basis corresponding to the mass-eigenstate basis of charged leptons.\footnote{The one-loop relation can be found in~\cite{Buttazzo:2020eyl}.}
Saturating the experimental results  leads to 
\begin{align}
\text{Re}\  \CL{e\gamma}{22}^\prime \approx  1.0 \times 10^{-5} \, \mathrm{TeV}^{-2} \, .
\label{eq:bound_C_egamma_22}
\end{align}

The tree-level expression of a generic radiative LFV rate in terms of the $\cC_{e\gamma}$ coefficients is 
\be
\mathcal{B}(\ell_i \to \ell_j \gamma) = \frac{m_{\ell_i}^3 v^2}{8 \pi \Gamma_{\ell_i}} \left(|\CL{e\gamma}{ij}'|^2 + |\CL{e\gamma}{ji}'|^2\right) \, .
\label{eq:Branching-ratio_lepton-decay}
\ee
Using this expression, the experimental bound $\mathcal{B}\!\brackets{\mu^+ \to e^+ \gamma} < 4.2 \times 10^{-13}$~(90\%~C.L.)
 obtained by the MEG experiment~\cite{MEG:2016leq} can be translated into the upper bound
\be
|\CL{e\gamma}{12(21)}^\prime| <  2.1 \times 10^{-10} \, \mathrm{TeV}^{-2} \, .
\label{eq:bound_C_egamma_12}
\ee
Taking into account Eq.~(\ref{eq:bound_C_egamma_22}), the requirement of fitting the  $a_\mu$ anomaly and, at the same time,
being consistent with the $\mathcal{B}\!\brackets{\mu^+ \to e^+ \gamma}$ bound, 
leads to the following tight constraints on off-diagonal over diagonal entries in the $2\times 2$ light lepton sector:
\be
 |\epsilon^L_{12}|\,, ~ |\epsilon^R_{12}|  <  2\times 10^{-5}\,,
  \label{eq:bound12}
 \ee
 where we have defined
 \be
\epsilon^{L(R)}_{12} \equiv \frac{ \CL{e\gamma}{12(21)}^\prime  }{    \CL{e\gamma}{22}^\prime    }\,, 
\qquad\qquad 
\epsilon^{L(R)}_{23} \equiv
   \frac{ \CL{e\gamma}{23(32)}^\prime  }{    \CL{e\gamma}{33}^\prime    }\,.
     \label{eq:epsdef}
\ee
The parameters $\epsilon^{L(R)}_{23}$ can be constrained by the bounds on radiative LFV decays of the $\tau$~lepton.
In particular,  $\mathcal{B}\!\brackets{\tau^\pm \to \mu^\pm \gamma} < 4.4 \times 10^{-8}$~(90\%~CL) as measured by the BaBar experiment~\cite{BaBar:2009hkt} implies
\begin{align}
|\CL{e\gamma}{23(32)}^\prime|  < 2.7 \times 10^{-6} \, \mathrm{TeV}^{-2} \,
\end{align}
that, in turn, leads to
\be
|\epsilon^{L}_{23} |\,, ~ |\epsilon^R_{23}| <   1.6 \times 10^{-2} \times 
 \left|  \frac{ y_\tau\, \CL{e\gamma}{22}^\prime  }{  y_\mu\,   \CL{e\gamma}{33}^\prime    }\right|\,.
 \label{eq:bound32}
\ee
In absence of a direct experimental constraint on the anomalous magnetic moment of the $\tau$~lepton, 
the normalisation of the bounds in Eq.~(\ref{eq:bound32}) has been chosen following the 
natural expectation 
\be
|   \CL{e\gamma}{33}^\prime|  /  y_\tau     \sim  |    \CL{e\gamma}{22}^\prime| / y_\mu ~. 
\ee

\section{RG evolution of the leptonic dipoles in the SMEFT}
\label{sect:RG}

In this section we analyse how the low-energy constraints derived before translate into 
high-scale constraints. To this purpose, we consider all possible $d=6$ operators with the same leptonic 
flavour structure, i.e.~operators of the type
\be
 \overline{\ell}_r \Gamma (A,H,\psi) e_s~, \qquad     \ell_r \equiv \begin{pmatrix} \nu_{L_r} \\ e_{L_r}  \end{pmatrix}~,   \quad e_s \equiv e_{R_s}~. 
 \label{eq:ops_gen}
 \ee
Those operators undergo a non-trivial mixing together with the dipole operators  and/or the Yukawa couplings. 
On the other hand, we can safely ignore operators with a different flavour structure since either they do not mix with 
dipole (or Yukawa) operators or they 
provide only a trivial multiplicative renormalisation.  

Adopting the SMEFT Warsaw basis~\cite{Grzadkowski:2010es} for the $d=6$ effective operators, 
the list of relevant terms can be decomposed as 
\be
\Delta \cL_\mathrm{unbroken}  =   \Delta \cL_H + \Delta \cL_{4f} + \mathrm{h.c.}  \,,
\label{ew:DLunbroken}
\ee
where 
\bea
 \Delta \cL_H  &=& - [Y_e]_{pr} (\overline{\ell}_p e_r) H + \C{eH}{pr} (\overline {\ell}_p e_r) H (H^\dagger H)  \no\\
&&+ \C{eB}{pr} (\overline{\ell}_p \sigma^{\mu\nu} e_r) H B_{\mu\nu} + \C{eW}{pr} (\overline{\ell}_p \sigma^{\mu\nu} e_r) \tau^I H W_{\mu\nu}^{I} \,,
\\
\Delta \cL_{4f} &=& 
 \C{lequ}{prst}^{(3)} (\overline{\ell}_p^j \sigma_{\mu\nu} e_r) \epsilon_{jk} (\overline{q}_s^k \sigma^{\mu\nu} u_t) 
+ \C{lequ}{prst}^{(1)} (\overline{\ell}_p^j e_r) \epsilon_{jk} (\overline{q}_s^k u_t)
\no \\
&&+ \C{ledq}{prst} (\overline{\ell}_p^j e_r) (\overline{d}_s q_{tj}) \,.
\label{eq:ourSMEFT}
\eea
In order to identify the mass-eigenstate basis for the leptons, and  the dipole operators defined in (\ref{eq:dipoledef}),
we need to work in the broken phase of the SMEFT. 
To this purpose, we can rewrite $\Delta \cL_H$ in the broken phase  as
\bea
 \Delta \cL^\mathrm{broken}_H  &=& 
- \squarebrackets{\mathcal{Y}_e}_{pr} \frac{v}{\sqrt 2} (\bar e_{Lp}e_{Rr} )
- \squarebrackets{\mathcal{Y}_{he}}_{pr} \frac{h}{\sqrt{2}} (\bar e_{Lp}e_{Rr} )  +\CL{e\gamma}{pr} \frac{v}{\sqrt 2} (\bar e_{Lp} \sigma^{\mu\nu} e_{Rr})F_{\mu\nu} 
\no \\
&& 
+ \CL{eZ}{pr} \frac{v}{\sqrt 2} (\bar e_{Lp} \sigma^{\mu\nu} e_{Rr})Z_{\mu\nu} + \mathcal{O}(h^2, h F_{\mu\nu} , h Z_{\mu\nu}) \,,
\eea
where $Z_{\mu\nu}$ is the field strength tensor for the $Z$ boson and $h$ is the physical Higgs boson.
The relations between terms in the broken and unbroken phase are 
\begin{align}
\label{eq:mass_Yukawa_broken_phase}
\begin{pmatrix}
	\CL{e\gamma}{rs} \\ \CL{eZ}{rs}
\end{pmatrix}
&= \begin{pmatrix}
	c_\theta & -s_\theta \\
	-s_\theta & -c_\theta
\end{pmatrix}
\begin{pmatrix}
	\C{eB}{rs} \\ 
	\C{eW}{rs}
\end{pmatrix} \,, 
\\[0.2cm]
\begin{pmatrix}
	[\mathcal{Y}_e]_{rs} \\
	[\mathcal{Y}_{he}]_{rs}
\end{pmatrix} 
&=
\begin{pmatrix}
	1 & -\frac{1}{2} \\
	1 & -\frac{3}{2}
\end{pmatrix}
\begin{pmatrix}
	[Y_e]_{rs} \\
	v^2 \C{eH}{rs}
\end{pmatrix} \, ,
\end{align}
where
\begin{equation}
	c_\theta = \frac{g_2}{\sqrt{g_1^2 + g_2^2}} = \frac{e}{g_1} \,, \qquad
 s_\theta = \frac{g_1}{\sqrt{g_1^2 + g_2^2}} = \frac{e}{g_2}
\end{equation}
with $g_2$, $g_1$, and $e$, the coupling constants of $SU(2)_L$, $U(1)_Y$, and $U(1)_{\rm em}$ respectively.


\subsection{Renormalization group equations}
The RG equations  for the SMEFT Wilson coefficients can be written in the form
\begin{align}
\mu \frac{\dd}{\dd \mu} C_i &= \frac{\dd}{\dd\log\mu} C_i = \frac{1}{16\pi^2} \gamma_{ij} C_j = \frac{1}{16\pi^2} \beta_i \,,
\label{eq:RGE}
\end{align}
where the $\gamma_{ij}$ (and $\beta_i$)  in the Warsaw basis have been derived in \cite{Jenkins:2013zja,Jenkins:2013wua,Alonso:2013hga} and are summarised in Ref.~\cite{Celis:2017hod}. We follow the Higgs potential normalisation conventions and the hypercharge assignments of the latter reference. In the limit in which we set all Yukawa couplings to zero, except for the top and bottom Yukawa~$[Y_{u(d)}]_{33}=y_{t (b)}$, the $\beta_i$ relevant to our analysis are

\begin{align}
\beta_{\underset{rs}{Y_e}} =&~ 2m^2 \squarebrackets{3 \C{eH}{rs} + N_c y_b \C{ledq}{rs33} - N_c y_t^\ast \C{lequ}{rs33}^{(1)}} \, , 
\label{eq:beta-Ye}
\\[0.1cm]
\begin{split}
\beta_{\underset{rs}{eH}} =&  \left[ 12 \lambda + 3 N_c \brackets{\abs{y_t}^2 + \abs{y_b}^2}  -  3 \brackets{3\mathsf{y}_l^2 + 3\mathsf{y}_e^2 - 4\mathsf{y}_l \mathsf{y}_e} g_1^2 + \frac{27}{4} g_2^2 \right] \C{eH}{rs}   \\
& - 6 \squarebrackets{4 g_1^3 \mathsf{y}_h^2 \brackets{\mathsf{y}_e + \mathsf{y}_l} + g_2^2 g_1 \mathsf{y}_h} \C{eB}{rs} \\
& - 3 \squarebrackets{4 g_1^2 g_2 \mathsf{y}_h \brackets{\mathsf{y}_e + \mathsf{y}_l} + 3 g_2^3} \C{eW}{rs} \\
\label{eq:beta-CeH}
&  + 2 N_c y_b \brackets{\lambda - 2 \abs{y_b}^2} \C{ledq}{rs33} - 2 N_c y_t^\ast \brackets{\lambda - 2 \abs{y_t}^2} \C{lequ}{rs33}^{(1)} \,,
\end{split}
\\[0.1cm]
\begin{split}
\beta_{\underset{rs}{eW}}=&  N_c \left[ \brackets{\abs{y_t}^2 + \abs{y_b}^2} \C{eW}{rs} - 2 g_2 y_t^\ast \C{lequ}{rs33}^{(3)} \right. \\
& \left.+ \squarebrackets{\brackets{3 c_{F,2} - b_{0,2}} g_2^2 + \brackets{-3\mathsf{y}_e^2 + 8 \mathsf{y}_e \mathsf{y}_l - 3\mathsf{y}_l^2} g_1^2} \C{eW}{rs} + g_1 g_2 \brackets{3 \mathsf{y}_l - \mathsf{y}_e} \C{eB}{rs} \right] \, ,
\label{eq:beta-CeW}
\end{split}
\\[0.1cm]
\begin{split}
\beta_{\underset{rs}{eB}}=&  N_c \left[ \brackets{\abs{y_t}^2 + \abs{y_b}^2} \C{eB}{rs} + 4 g_1 \brackets{\mathsf{y}_u + \mathsf{y}_q} y_t^\ast \C{lequ}{rs33}^{(3)} \right. \\
 & \left. + \squarebrackets{-3 c_{F,2} g_2^2 + \brackets{3\mathsf{y}_e^2 + 4 \mathsf{y}_e \mathsf{y}_l + 3\mathsf{y}_l^2 - b_{0,1}} g_1^2} \C{eB}{rs} + 4 c_{F,2} g_1 g_2 \brackets{3 \mathsf{y}_l - \mathsf{y}_e} \C{eW}{rs} \right] \, ,
\label{eq:beta-CeB}
\end{split}
\end{align}
where $c_{F,2}=3/4$, $b_{0,1}=-1/6-20n_g/9$, and $b_{0,2}=43/6-4n_g/3$ with the number of generations~$n_g$.

In terms of these expressions, working  at one-loop accuracy, the solutions to the RG equations of the electromagnetic dipole operators and the mass Yukawa in the broken phase assume the form
\begin{align}
\CL{e\gamma}{rs} (\mu_L) &= \CL{e\gamma}{rs} (\mu_H) + \frac{1}{16\pi^2} \log\brackets{\frac{\mu_L}{\mu_H}} \left( c_\theta \beta_{\underset{rs}{eB}} -s_\theta  \beta_{\underset{rs}{eW}} \right)~,  \\
[\mathcal{Y}_{e}]_{rs} \, (\mu_L) &= 
[\mathcal{Y}_{e}]_{rs} \, (\mu_H) + \loopfactor 
\left( \beta_{\underset{rs}{Y_e}}-\frac{v^2}{2} \beta_{\underset{rs}{eH}} \right) \,,
\end{align}
where we evolve the coefficients from a low scale~$\mu_L$ to the high scale~$\mu_H$.
Further neglecting the terms proportional to the quartic Higgs coupling or 
to at least two powers of the gauge couplings, assuming $y_t$ to be real, and defining $\hat{L} = \dfrac{1}{16\pi^2} \log \left( \dfrac{\mu_H}{\mu_L}\right)$,
leads to
\begin{align}
\CL{e\gamma}{ij} (\mu_L) 
&= \squarebrackets{1 - 3 \hat{L}  \brackets{y_t^2  + y_b^2}} \CL{e\gamma}{ij} (\mu_H) - \squarebrackets{16\hat{L} y_t e} \C{lequ}{ij33}^{(3)} (\mu_H) \,,\\
[\mathcal{Y}_e]_{ij} (\mu_L) &=  \squarebrackets{Y_{e}}_{ij}(\mu_H) - \frac{v^2}{2}\C{eH}{ij}(\mu_H) + 6  v^2 \hat{L}
\left[ y_t^3 \C{lequ}{ij33}^{(1)}- y^3_b \C{ledq}{ij33} +  \frac34 (y_t^2  + y_b^2) \C{eH}{ij} \right]_{\mu_H}~. \quad
\label{eq:physical_Yukawa_low}
\end{align}

\subsection{Rotation to the mass basis}
Focusing on the $2\times 2$ sector of light lepton indices, we define the following ratios of Wilson coefficients
at the high scale:
\begin{align}
\theta_{L}^Y  &= \left. \frac{ [Y_e]_{12}  }{   [Y_e]_{22} }  \right|_{\mu_H}   \,, &
\theta_{L}^{e\gamma}  &=  \left. \frac{ \CL{e\gamma}{12}   }{  \CL{e\gamma}{22}   } \right|_{\mu_H}    \,, &
\theta_{L}^{eH }  &=  \left. \frac{ \C{eH}{12}  }{ \C{eH}{22}  }  \right|_{\mu_H}  \,,  &
\theta_{L}^{u_i}  &= \left. \frac{ \C{lequ}{1233}^{(i)}   }{ \C{lequ}{2233}^{(i)}  }\right|_{\mu_H}   \,,  &
\theta_L^{d}   &=   \left.  \frac{ \C{ledq}{1233}  }{ \C{ledq}{2233}  } \right|_{\mu_H}  \,. 
\label{eq:thetaX}
\end{align}
We denote these parameters the  left-handed \textit{flavour phases} of the operators.
In a similar fashion we define the right-handed flavour phases $\theta_{R}^X$ from the ratios of 21 over 22 entries on a given Wilson coefficient.
These  flavour phases define the alignment in flavour  space of these five operators. 
We assume $|\theta_{L(R)}^X| \ll 1$, and ever smaller values for the ratios of 11 over 22 entries (for each Wilson coefficient).

We are interested in estimating the size of the (flavour) off-diagonal entries of the dipole operator
in the flavour basis where the mass Yukawa at the low scale is diagonal.
In the limit of small off-diagonal entries, the rotation to the mass basis in the 12 sector is determined by 
\begin{align}
\Theta^{\mathcal{Y}}_{L(R)} &= - \left. \dfrac{[\mathcal{Y}_{e}]_{12(21)} }{[\mathcal{Y}_{e}]_{22} }\right|_{\mu_L}
\end{align}
and, in general, we expect  $\Theta^{\mathcal{Y}}_{L(R)}\neq \theta_{L(R)}^Y$.
In the mass basis 
the off-diagonal elements of the dipole operator are given by
\be
	 \CL{e\gamma}{12(21)}'(\mu_L) = \CL{e\gamma}{12(21)} (\mu_L) + \Theta_{L(R)}^{ \mathcal{Y}   } ~\CL{e\gamma}{22} (\mu_L)
	 \label{eq:dipole_mass-basis-rotation} 
\ee
while the diagonal element receives negligible corrections i.e.~$\CL{e\gamma}{22}'(\mu_L) \approx \CL{e\gamma}{22} (\mu_L)$.

Focusing on the 12 sector, i.e.~on left-handed flavour rotations, and expressing the $\cC_{e\gamma}$ and  
$\mathcal{Y}_{e}$ coefficients in terms of their low-scale values, we obtain
\bea
\CL{e\gamma}{12}'(\mu_L) &=&  (\theta_{L}^{e\gamma}-\theta_{L}^Y) \CL{e\gamma}{22}  (\mu_L) +    (\theta_{L}^{e\gamma}-\theta_{L}^{u_3}) (16 \hat{L} e y_t)  \C{lequ}{2233}^{(3)} (\mu_H)  \no\\
&+&  \left[   (\theta_{L}^Y-\theta_{L}^{u_1})  (6 y_t^3 ) \C{lequ}{2233}^{(1)}  (\mu_H)
+ (\theta_{L}^{d}-\theta_{L}^Y)  (6 y_b^3 ) \C{ledq}{2233} (\mu_H) \right]   \frac{ 1 }{  [\mathcal{Y}_{e}]_{22} (\mu_L)}  \hat{L}  v^2  \CL{e\gamma}{22} (\mu_L)
   \no \\ 
&+&   (\theta_{L}^{eH } -\theta_{L}^Y)    \frac{1 -9( y_t^2 +y_b^2) \hat{L}  }{2} \C{eH}{22}  (\mu_H)   \frac{ 1 }{  [\mathcal{Y}_{e}]_{22}  (\mu_L)}   v^2  \CL{e\gamma}{22} (\mu_L)~. 
\qquad 
\label{eq:Ceg-master}
\eea
The above equation allows us to derive a few important considerations:
\begin{itemize}
\item
Even if $\CL{e\gamma}{12} =0$ at the high scale (i.e.~if~$\theta_L^{e\gamma}=0$), a non-vanishing 
$\CL{e\gamma}{12}'(\mu_L)$ is naturally generated at the low scale if {\em any} of the other operators
in (\ref{ew:DLunbroken}) has a non-vanishing 12 entry (i.e.~if any of the $\theta_L^{X}$ in Eq.~(\ref{eq:thetaX}) is non-zero).
\item 
$\CL{e\gamma}{12}'(\mu_L)=0$ is obtained only aligning {\em all} the flavour phases in Eq.~(\ref{eq:thetaX}).
\item 
All terms but one in Eq.~(\ref{eq:Ceg-master}) are proportional to the 22 entry of the dipole operator, which is required to be non-vanishing and 
sizeable by the $a_\mu$ anomaly (see Sect.~\ref{sect:exp}).  
\end{itemize}
Of the terms in Eq.~(\ref{eq:Ceg-master}), the one proportional to $C_{ledq}$ has a numerically small coefficient which does not imply a sizeable 
tuning on the  difference $(\theta_{L}^{d}-\theta_{L}^Y)$.  The combination 
$(\theta_{L}^{eH } -\theta_{L}^Y)  C_{eH}$ controls the $e$--$\mu$  flavour violating coupling of the physical Higgs boson, which is tightly constrained 
by other observables~\cite{Blankenburg:2012ex,Harnik:2012pb}  and can be safely ignored in the present analysis. The non-trivial contributions to 
$\epsilon^L_{12}$ defined in Eq.~(\ref{eq:epsdef}) can be thus simplified to 
\bea
\epsilon^L_{12} =  (\theta_{L}^{e\gamma}-\theta_{L}^Y)  +
  (\theta_{L}^{u_3}- \theta_{L}^{e\gamma})  \Delta_{3}  + (\theta_{L}^{u_1} - \theta_{L}^Y)   \Delta_{1} 
  \label{eq:main12}
\eea
with 
\bea \label{eq:Delta3}
\Delta_{3}  &=&     \frac{- 16 \hat{L} e y_t   \C{lequ}{2233}^{(3)} (\mu_H) }{    \CL{e\gamma}{22} (\mu_L)  }
=    \frac{  \left. \Delta a_\mu  \right|_{ C_{lequ}^{(3)} }   }{    \Delta a_\mu^{\rm exp}   } 
\label{eq:D3} \\
\Delta_{1}  &=&   \frac{ - 6 y_t^3  \hat{L}  v^2 }{  [\mathcal{Y}_{e}]_{22}  (\mu_L)}  \C{lequ}{2233}^{(1)} (\mu_H)   \approx 1 \times 10^{-3} \left[  \frac{   \C{lequ}{2233}^{(1)} (\mu_H)  }{    \C{lequ}{2233}^{(3)} (\mu_H)  }    \right] \times \Delta_3 \,.
\label{eq:D1}
\eea
By construction, the coefficient $\Delta_{3}$ is of $\mathcal{O}(1)$
if $\Delta a^{\rm exp}_\mu$ is saturated by the (radiatively induced) 
contribution of the triplet semileptonic operator. 
In this limit, assuming further triplet and singlet semileptonic operators 
of similar size,  we expect  $\Delta_{1}= \mathcal{O}(10^{-3})$.
In the case where the muon Yukawa is saturated by the (radiatively induced) contribution of the singlet semileptonic operator we expect $\Delta_1$ of $\mathcal{O}(1)$.

Given the $10^{-5}$ bound on $|\epsilon^L_{12}|$ in Eq.~(\ref{eq:bound12}), 
the result in Eq.~(\ref{eq:main12}) implies a tight alignment in flavour space
of four a priori independent SMEFT  operators. Note that the bound is expressed in terms of the left-handed flavour phases  ($\theta^X_L$), 
which characterise the orientation of the operators in flavour space, and not in terms of the size of the Wilson coefficients.
In the case of $C_{lequ}^{(1,3)}$, the impact of the size is encoded by the coefficients 
$\Delta_{1,3}$ which, as shown in Eqs.~(\ref{eq:D3})--(\ref{eq:D1}), are expected to be sizeable. 

A completely analogous formula holds for the  $23$ sector; however, in that case we can 
neglect the term proportional to $\Delta_1$, whose size is small enough (in case of similar size triplet and singlet semileptonic operators) to 
satisfy the constraint on $|\epsilon^L_{23}|$ even in presence of $\mathcal{O}(1)$  
misalignments in flavour space.

\subsection{Light NP contributions to $a_\mu$}

As stated in the introduction, our analysis is focused on NP models well described by the SMEFT. However, it is worth to point out that 
there are also explanations of the muon anomalous magnetic moment based on {\em light} new physics, i.e.~in terms of new 
degrees of freedom with $m_{\rm NP} < v$ (see e.g.~\cite{Marciano:2016yhf,Bauer:2019gfk,Cornella:2019uxs,Buen-Abad:2021fwq,Brdar:2021pla}).

For light NP  the last two terms in Eq.~\eqref{eq:main12} are absent since no significant RG contributions are expected below 
the top quark mass scale ($m_t \sim v$).
On the other hand, the first term in Eq.~\eqref{eq:main12} still needs to be aligned since also 
light NP contributions to $a_\mu$ are subject to the strong flavour alignment conditions 
in Eq.~\eqref{eq:bound12} and~\eqref{eq:bound32}. 
The dipole operator flavour phase $\theta_L^{e\gamma}$ should be interpreted as the flavour phase of the effective photon di-lepton coupling (including short- and long-distance contributions) while the lepton Yukawa flavour phase $\theta_L^Y$ is the angle in the charged lepton mass matrix.
The flavour-alignment problem posed by $\mu \to e \gamma$ on NP contributions to $a_\mu$ is therefore a serious constraint also for the 
light NP explanations.

Moreover, LFV couplings of light NP can mediate additional LFV processes which are disconnected from the operators of our heavy NP analysis. For example, in the case of an axion-like particle explanation of $\Delta a_\mu$, other processes such as $\mu \to e + \text{invisible}$ and $\mu \to e\bar ee$ provide additional
constraints on the LFV couplings of the underlying model~\cite{Bauer:2019gfk}.

\section{General alignment mechanisms}
\label{sect:align}

Barring accidental cancellations, the alignment in Eq.~(\ref{eq:main12})  can be realised in general terms 
following two main strategies: either via dynamical assumptions, or 
by imposing appropriate flavour symmetries. 
In this section we illustrate these two possibilities in general terms, while in Sect.~\ref{sect:mediators} 
we discuss their explicit implementation in a specific NP model with scalar mediators. 

\subsection{Dynamical conditions}
\label{sect:dynamical}

The alignment via dynamical assumptions occurs if the UV dynamics is 
such that two or more SMEFT operators originate from the same short-distance structure 
and hence appear with the same orientation in flavour space.  
We should distinguish two main cases: i)~the case $\Delta_3 = \mathcal{O}(1)$,
when the NP contribution $\Delta a_\mu$ is dominated by the term induced by $\cC_{lequ}^{(3)}$,  and ii)~the case $|\Delta_{1,3}| \ll 1$.

\begin{table}[t]
\begin{center}
\begin{tabular}{c|rc|c}
 Case &  & Dynamical hypothesis & Alignment condition \\ \hline\hline
  & {\bf I})  & Dipole operator radiatively generated with $C_{lequ}^{(3)}$ &  $\theta_{L}^{e\gamma} = \theta_{L}^{u_3}$ \\
 $\Delta_3 = \mathcal{O}(1)$  & {\bf II}) &  $C_{lequ}^{(1)}$ and $C_{lequ}^{(3)}$ from same UV dynamics &  $\theta_{L}^{u_1} = \theta_{L}^{u_3}$ \\
& {\bf III})  & $y_\mu$ radiatively generated with $C_{lequ}^{(1)}$  & $\theta_{L}^Y= \theta_{L}^{u_1}$  \\[2pt] \hline\hline
\multirow{2}{*}{ $|\Delta_{1,3}| \ll 1$} 
&  &  Dipole operator  and lepton Yukawa  &  
\multirow{2}{*}{  $\theta_{L}^{e\gamma} =\theta_{L}^Y$}   \\[-3pt]
&  &  radiatively generated  by  same UV dynamics &      \\ \hline\hline
\end{tabular}
\caption{Alignment conditions following from specific dynamical assumptions.
\label{tab:dynamical}}
\end{center}
\end{table}

\paragraph{i) $\Delta_3 = \mathcal{O}(1)$.}
As shown in Table~\ref{tab:dynamical},  in this  case we can identify three dynamical hypotheses which lead to specific alignments in flavour space  in a wide class of explicit models.  If {\em all} the three conditions in Table~\ref{tab:dynamical} 
are satisfied we do not have a severe alignment problem in Eq.~(\ref{eq:main12}).

The condition {\bf I} in Table~\ref{tab:dynamical} is the most natural one: it consists in assuming that the dipole operator is not present in the 
UV theory, but is radiatively generated by the same dynamics giving rise to $C_{lequ}^{(3)}$
 (e.g.~via the tree-level exchange of a leptoquark~(LQ) field). In this case we clearly 
 have $\theta_{L}^{e\gamma} = \theta_{L}^{u_3}$. 
Note that  $\Delta_3$ is not necessarily one, since there can be finite UV matching contributions to $\cC_{e\gamma}$; however finite contributions and RG induced terms 
are, by hypothesis,  flavour aligned.
 The condition {\bf II} is already more restrictive: it implies that two semileptonic operators with different 
 electroweak structures originate from the same underlying UV dynamics. 
 More precisely, a unique combination of flavour symmetry breaking terms controls
 the orientation of the two operators in flavour space.
 Such condition can be realised assuming 
 e.g.~both  $C_{lequ}^{(1)}$ and  $C_{lequ}^{(3)}$
 arise by the tree-level exchange of a single LQ field (as we discuss explicitly 
 in Sect.~\ref{sect:mediators}), but we must ensure they do not receive additional 
 contributions (e.g.~a heavy colourless scalar field would contribute to $C_{lequ}^{(1)}$ but not to $C_{lequ}^{(3)}$).

More challenging to realise is the condition {\bf III}, which requires that not only the dipole
operator is radiatively generated, but also the Yukawa coupling. This is more difficult to conceive since the 
Yukawa operator is a marginal operator ($d=4$) which is naturally present in the theory at all scales,
at least as long as the corresponding fields are well defined. 
One can forbid the Yukawa coupling at the tree-level assuming extra symmetries that, however,
must be broken in other sectors of the theory.
This should be 
contrasted to the $d=6$ dipole operator, that is naturally absent in a well-behaved UV completion. 
A further complication arises by the fact that $C_{lequ}^{(1)}$ corresponds to a rank-one
tensor in lepton space once we consider only the third-generation quarks
(and trace over the quark-flavour indices). One can therefore generate only part of the full
Yukawa coupling via $C_{lequ}^{(1)}$, ideally the leading entries in the $2\times2$ 
sub-block corresponding to the light families. The remaining terms need to be generated 
by additional operators (e.g.~via semilpetonic operators involving light quark families or
exotic fermions), and the tuning problem is simply shifted to a different set of operators
in the effective theory. 
Because of the difficulties associated to the condition {\bf III},  the possibility of addressing the alignment problem in Eq.~(\ref{eq:main12}) 
using {\em only} dynamical hypotheses appears to be rather unnatural.

\paragraph{ii) $|\Delta_{1,3}| \ll  \mathcal{O}(1)$.}
If $\Delta_{1,3}$ are sufficiently small, the only condition 
we need to care about in order to satisfy Eq.~(\ref{eq:main12}) is the flavour
alignment of dipole operator and lepton Yukawa coupling. First, it is worth noting that 
we enter into this regime only if $|\Delta_{3}| \lsim 10^{-5}$ that, by itself, is a rather
unnatural condition. Second, we observe that in this case the problem is a genuine 
UV boundary condition in the SMEFT and is not related to the RG structure.
This condition can be realised dynamically if dipole operator  and lepton Yukawa
are generated by the same UV dynamics (beyond the SMEFT). Models of this type have been proposed for instance in Ref.~\cite{Baker:2021yli,Yin:2021yqy,Greljo:2021npi} (see also Ref.~\cite{Gabrielli:2013jka,Gabrielli:2016cut}). 
As already commented for the case i), the fact that 
dipole operators  and lepton Yukawa have different canonical dimension makes 
the flavour alignment in these constructions rather unnatural, 
unless supplemented by  additional flavour symmetries (as e.g.~in Ref.~\cite{Greljo:2021npi}).

\subsection{Flavour symmetries}
\label{sect:flavsymm}

Exact or approximate alignments in flavour space can be achieved by means of exact or approximate 
global symmetries. Here we discuss two representative cases:  the individual lepton numbers, and the 
$U(2)_{L_L} \times U(2)_{E_R}$ flavour symmetry acting on the light lepton families.

\subsubsection{$U(1)$ symmetries}
The three individual lepton numbers, 
\be
U(1)_{L_e} \times U(1)_{L_\mu} \times U(1)_{L_\tau}  = 
U(1)_{L} \times  U(1)_{L_e - L_\mu} \times  U(1)_{L_\tau-L_\mu} 
\label{eq:U1}
\ee
are, separately, exact accidental symmetries of the $d=4$ operators in the SMEFT.  
As shown on the r.h.s.~of Eq.~(\ref{eq:U1}), we can rearrange these symmetries 
into total lepton number  ($L= L_e+ L_\mu+ L_\tau$), and two $L$--conserving 
Abelian groups. Assuming the latter to be a good symmetry of the UV theory implies that  
all the flavour phases in (\ref{eq:thetaX}) are zero. The symmetry (\ref{eq:U1}) must be 
broken in the neutrino sector; however, the smallness of neutrino masses and their Majorana 
nature allows us to assume that the three individual lepton numbers are 
conserved to a high accuracy in all the operators preserving 
total lepton number. 

An almost exact conservation of the individual lepton numbers is the assumption employed 
in many recent explicit models proposed to address the $(g-2)_\mu$ anomaly
(see in particular~\cite{Greljo:2021xmg,Davighi:2021oel,Arcadi:2021cwg,Greljo:2021npi,Cen:2021iwv}).
Note in particular that any combination of  $U(1)_{L_\mu}$ and $U(1)_{L_\tau}$
 is sufficient to protect the tightly constrained $e$--$\mu$ mixing
for heavy NP as well as for light NP.

\subsubsection{$U(2)_{L_L} \times U(2)_{E_R}$}
The $U(2)$ flavour symmetries acting on the light families of each SM field, with small breaking terms, 
are introduced with a twofold purpose~\cite{Barbieri:2011ci,Barbieri:2012uh,Blankenburg:2012nx}: 
i) explaining the hierarchical structure of the SM Yukawa couplings
(only the third generation couplings are allowed by the symmetry); 
ii) allowing TeV-scale NP coupled to the third generation,  possibly addressing 
the Higgs hierarchy problem, in a way that is consistent with the tight bounds from flavour violating 
processes (which necessarily involve the light families, and hence are suppressed by the symmetry).
As proposed in Ref.~\cite{Greljo:2015mma,Barbieri:2015yvd} 
(see also~\cite{Bordone:2017anc,Fuentes-Martin:2019mun,Cornella:2021sby})
this approach turns out to be very successful for a combined explanation of 
both charged- and neutral-current $B$~physics anomalies.

Focusing the attention to the lepton sector, the relevant symmetry is $U(2)_{L_L} \times U(2)_{E_R}$.
This is assumed to be broken only by two spurions, $V_\ell = (2,1)$ and
$\Delta_e = (2,\bar{2})$, such that the charged-lepton Yukawa coupling assumes the form 
\begin{align}
	Y_e &= y_\tau \begin{pmatrix}
		\Delta_e &   V_\ell \\
		0 & 1
\end{pmatrix}~.
\end{align}
The two spurions can be parametrised by
\begin{align}
	V_{\ell} &= \begin{pmatrix}
		0 \\ \epsilon_{\ell}
	\end{pmatrix} \, , & 
	\Delta_e &= O_e^\intercal \begin{pmatrix}
		\delta^\prime_e & 0 \\
		0 & \delta_e
	\end{pmatrix}
	\label{eq:standardbasis}
\end{align} 
with  $|\delta_e^\prime|  \ll  |\delta_e|  \ll  |\epsilon_\ell| \ll 1$ and 
where $O_e$ is a real orthogonal matrix ($[O_e]_{12} = \sin \theta_e \equiv s_e$).
It must be stressed that, while we can deduce the size of $\delta_e$ and $\delta_e^\prime$
by the eigenvalues of the charged-lepton Yukawa couplings, the size of $\epsilon_\ell$ and $s_e$ cannot be 
directly deduced from~$Y_e$. Here we assume $\epsilon_\ell =\mathcal{O}(10^{-1} )$, which is the most natural 
choice for a similar treatment of quark and lepton sectors~\cite{Blankenburg:2012nx},
and is also the value supported by the recent data on the
 $B$~physics anomalies~\cite{Greljo:2015mma,Barbieri:2015yvd}. Similarly, we 
 expect  $s_e=\ord{ \sqrt{m_e/m_\mu}} \gsim 10^{-2}$~\cite{Bordone:2017anc,Fuentes-Martin:2019mun}.

The flavour orientation of all the effective operators relevant to our analysis 
is determined,  up to $\ord{1}$ coefficients,
by the small breaking terms appearing in Eq.~(\ref{eq:standardbasis}).
In practice, this characterisation is achieved constructing the effective operators via a perturbative
expansion in terms of the spurions~\cite{Faroughy:2020ina},
requiring the theory to be invariant 
under $U(2)_{L_L} \times U(2)_{E_R}$.

\paragraph{$e - \mu$ sector}
Non-vanishing entries for operators of the type (\ref{eq:ops_gen})  in the 
12 sector appear at $\ord{\Delta_e}$ in the spurion expansion. 
At this order all the flavour phases are aligned (to the phase of $\Delta_e$) and the condition (\ref{eq:main12}) is
satisfied automatically. However, this is not the case to higher orders in the spurion expansion. 

At order $\ord{V_\ell^2 \Delta_e}$  we can express the combination 
of spurions controlling the flavour structure of any operator  of the type 
(\ref{eq:ops_gen}) as
\begin{align}
	X_{\alpha\beta}^n &= a_n (\Delta_e)_{\alpha\beta} + b_n (V_\ell)_\alpha (V_\ell^\dagger)_\gamma (\Delta_e)_{\gamma\beta} 
	= \begin{pmatrix}
		a_n c_e \delta_e^\prime & -a_n s_e \delta_e \\
		s_e \delta_e^\prime (a_n+b_n\epsilon_\ell^2) & c_e \delta_e (a_n+b_n\epsilon_\ell^2)
	\end{pmatrix}_{\alpha\beta} \,,
\end{align}
where $a_n$ and $b_n$ are $\ord{1}$ coefficients. 
The flavour matrix $X_{\alpha\beta}^n$ is defined such that the corresponding operator reads $X_{\alpha\beta}^n (\bar{\ell}_\alpha \Gamma e_\beta) \eta^n$, 
where $\eta^n$ denotes the lepton independent structure $n\in\{ Y,eH,e\gamma,u_3,u_1,d \}$.  
For each operator the $X^n$ are diagonalised by two unitary matrices, $O_{L,n}$ and $O_{R,n}$, 
whose rotation angles are 
\be
	\theta_L^n  \approx \frac{s_e}{c_e} \frac{1}{1+\frac{b_n}{a_n}\epsilon_\ell^2} \approx \frac{s_e}{c_e} \brackets{1 - \frac{b_n}{a_n}\epsilon_\ell^2} \, ,
	\qquad\
	\theta_R^n  \approx - \frac{s_e}{c_e} \frac{\delta_e^\prime}{\delta_e} \, .
	\label{eq:anbn}
\ee
Note that $\theta_R^n$ is independent of $n$, therefore all operators are 
aligned in the $U(2)_{E_R}$ space. This is a consequence of having assumed a single $U(2)_{E_R}$
breaking spurion. On the other hand,  misalignments are possible in the $U(2)_{L_L}$ space.
In such space, the difference in the flavour orientation between operators $m$ and $n$ is
controlled by
\begin{align}
	\Delta\theta_L^{nm} &= \theta_L^m - \theta_L^n = \frac{s_e}{c_e} \epsilon_\ell^2 \brackets{\frac{b_n}{a_n}-\frac{b_m}{a_m}} = \frac{s_e}{c_e} \epsilon_\ell^2 \brackets{d_n-d_m} \, ,
\end{align}
where $d_i \equiv b_i/a_i$. Looking in particular to the 
relative phase of the dipole operator and the Yukawa interaction, from 
the condition (\ref{eq:main12}) and the bound on  $|\epsilon^L_{12}|$ in (\ref{eq:bound12}) we obtain
\begin{align}
	\left|  \theta_{L}^{e\gamma}-\theta_{L}^Y  \right|   =\left| \frac{s_e}{c_e}  \right| \epsilon^2_\ell \left|d_Y - d_{e\gamma}\right| \leq 2 \times 10^{-5}.
\end{align}
Setting $c_e=\ord{1}$ and $\epsilon_\ell=\ord{10^{-1}}$~\cite{Faroughy:2020ina,Cornella:2021sby}, this leads to 
\begin{align}
	\abs{ s_e (d_Y - d_{e\gamma})} \lesssim 10^{-3}~.
\end{align}
Given the natural expectation  $s_e=\ord{ \sqrt{m_e/m_\mu}}$~\cite{Bordone:2017anc,Fuentes-Martin:2019mun},
this implies a tight alignment condition on the $\ord{1}$ coefficients~$d_i$.

\paragraph{$\mu - \tau$ sector}
We can proceed  in a similar way in the 23 sector. In this case off-diagonal coefficients in flavour space appear already at 
$\ord{V_\ell}$.  Denoting $Y_{\alpha\beta}^n$  the combination 
of spurions controlling the flavour structure of the operators in the 23 sector we find 
\begin{align}
	Y_{\alpha\beta}^n &= \begin{pmatrix}
		0 & f_n \epsilon_\ell \\
		0 & g_n
	\end{pmatrix} \,,
\end{align}
where $f_n,g_n=\ord{1}$. Defining $h_i \equiv f_i/g_i$ and proceeding as in the 12 sector, taking into account the 
 bound on  $|\epsilon^L_{23}|$ in (\ref{eq:bound32}), leads to
\be 
	\abs{ \epsilon_\ell (h_Y - h_{e\gamma})} \lesssim 2 \times 10^{-2}\quad \longrightarrow\quad   
	\abs{h_Y - h_{e\gamma}}\lesssim 2 \times 10^{-1}~.
\ee
Similarly to the 12 sector, also in this case we find a non-trivial constraint on coefficients which
are expected to be of~$\ord{1}$. 

In summary, while a minimally broken $U(2)_{L_L} \times U(2)_{E_R}$ symmetry does 
provide a partial alignment in the flavour space of the operators contributing to the dipole 
and Yukawa couplings at low energies,   this alignment is not enough to fully justify 
the smallness of $|\epsilon^L_{12}|$ and $|\epsilon^L_{23}|$.

\section{Alignment in an explicit NP model}
\label{sect:mediators}

In order to illustrate the general mechanism discussed in the previous section in concrete cases, 
we analyse here a simplified model where both semileptonic operators are generated by the tree-level 
exchange of scalar mediators. To this purpose, we note that we can get rid of the 
tensor currents present in Eq.~(\ref{eq:ourSMEFT}) changing basis from $Q_{lequ}^{(1,3)}$
to $Q_{S_1}$ and $Q_\Phi$, defined as 
\be
\left(    \begin{array}{l}  Q_{lequ}^{(1)} \\   Q_{lequ}^{(3)} \end{array} \right)
 = \left( \begin{array}{cc}   0  & 1 \\ 8 & 4 \end{array} \right) 
 \left(    \begin{array}{l}  Q_{S_1} \\   Q_{\Phi} \end{array} \right) \,,
 \qquad
\begin{array}{l}
Q_{S_1} = \epsilon_{jk} (\overline{\ell}^j  {q^c}^k ) (\overline{u}^c e) \,,  \\
Q_{\Phi} = \epsilon_{jk} (\overline{\ell}^j e) (\overline{q}^k u) \,.
\end{array}
\ee
This implies that a $S_1$ scalar leptoquark transforming as $(\bar{3},1)_{\frac13}$ under the SM gauge group, 
and 
a Higgs-like field $\Phi$ transforming as $(1,2)_{\frac12}$, allow us to generate generic 
tree-level matching conditions for  $Q_{lequ}^{(1,3)}$.  Interestingly enough, 
both these fields also lead to non-vanishing one-loop contributions to 
dipole amplitudes. 

The simplified renormalizable model we consider contains these two exotic scalar fields, 
coupled to the SM via the following Lagrangian
\bea
	\L_\mathrm{S_1} &=& \L_\mathrm{SM} + (D_\mu S_1)^\dagger (D^\mu S_1) - M_{S_1}^2 S_1^\dagger S_1 
	-\squarebrackets{\lambda_{i\alpha}^L (\bar{q}_i^c \epsilon \ell_\alpha)S_1 + \lambda_{i\alpha}^R (\bar{u}_i^c e_\alpha)S_1 + \mathrm{h.c.} } \qquad  \no\\
&&	\qquad +\ (D_\mu \Phi)^\dagger (D^\mu \Phi) - M_\Phi^2 \Phi^\dagger \Phi 
	- \squarebrackets{\lambda^e_{\alpha\beta} (\bar{\ell}_\alpha e_\beta) \Phi + \lambda^u_{ij} (\bar{q}_i u_j) \tilde{\Phi} + \mathrm{h.c.} } \,,
\label{eq:Lmodel}
\eea
where $\tilde{\Phi} \equiv \epsilon\, \Phi^*$.
Integrating out the $S_1$ and $\Phi$ fields at the tree-level leads to the following matching conditions for 
the Wilson coefficients of  $Q_{lequ}^{(1)}$ and  $Q_{lequ}^{(3)}$:
\begin{align}
	C_{\underset{\alpha\beta ij}{lequ}}^{(1)}  = 
            \frac{ {\lambda_{i\alpha}^L}^\ast \lambda_{j\beta}^R}{ 2 M_{S_1}^2 }  - 
            \frac{  \lambda^e_{\alpha\beta}\lambda^u_{ij}   }{ M_\Phi^2 }   ~, \qquad  
	C_{\underset{\alpha\beta ij}{lequ}}^{(3)}  = 
	 -   \frac{{\lambda_{i\alpha}^L}^\ast \lambda_{j\beta}^R}{8  M_{S_1}^2  }~. 
\end{align}
Given the interaction terms in (\ref{eq:Lmodel}),
this model also yields a direct contribution to the dipole operators 
from the integration of $S_1$ at the one-loop level.\footnote{In principle, also $\Phi$
can generate a one-loop contribution to dipole amplitudes. However, this arises only if we add 
a quartic interaction between $\Phi$ and the SM Higgs field, and this is formally described by  
dimension-8 operators in the SMEFT expansion.} The latter yields
\begin{align}
\begin{split}
	\CL{e\gamma}{\alpha\beta} (\mu_H)
	= \frac{e}{16 \pi^2 M_{S_1}^2} &\left\{ -\frac{1}{8} \squarebrackets{(\lambda^L)^\dagger \lambda^L Y_e}_{\alpha\beta} - \frac{1}{8}\squarebrackets{Y_e (\lambda^R)^\dagger \lambda^R}_{\alpha\beta} \right.
	\\
	&\quad + \left. \brackets{\frac{7}{4} + \log\frac{\mu_H^2}{M_{S_1}^2}} \squarebrackets{(\lambda^L)^\dagger Y_u^\ast \lambda^R}_{\alpha\beta} \right\}~,
	\label{eq:S1-dipole_contribution}
\end{split}
\end{align}
which agrees with the results in Refs.~\cite{Gherardi:2020det,Dorsner:2020aaz,
Dedes:2021abc,Marzocca:2021azj}.
We now have all the ingredients to determine the misalignment of the flavour phases entering 
Eq.~(\ref{eq:main12}) in this model, but for $\theta_L^Y$ which is a free parameter.
Setting $\mu_H=M_{S_1}$ we obtain the following expressions for the left-handed 
flavour phases in the 12 sector:
\bea
\theta_L^{u_1} &=& \frac{{\lambda^L_{31}}^\ast   \lambda_{32}^R  +2   \lambda^e_{12}\lambda^u_{33}  M_{S_1}^2 / M_{\Phi}^2  }{{\lambda^L_{32}}^\ast   \lambda_{32}^R  +2   \lambda^e_{22}\lambda^u_{33} M_{S_1}^2 / M_{\Phi}^2  } \,, 
	\\
 \theta_L^{u_3} &=&  \frac{{\lambda^L_{31}}^\ast}{{\lambda^L_{32}}^\ast} \,, 
	\\
	\theta_L^{e\gamma} &=& \frac{ (Y_e)_{1\alpha}  {\lambda^R_{i\alpha}}^\ast \lambda^R_{i2} +  {\lambda^L_{i1}}^\ast \lambda^L_{i \alpha} (Y_e)_{\alpha 2}
	-  14 \, y_t {\lambda^L_{31}}^\ast \lambda^R_{32}}{ (Y_e)_{2\alpha}   {\lambda^R_{i\alpha}}^\ast \lambda^R_{i2}   
	+  {\lambda^L_{i2}}^\ast \lambda^L_{i \alpha} (Y_e)_{\alpha 2}  
	 -  14\, y_t {\lambda^L_{32}}^\ast \lambda^R_{32}} \, ,
	\label{eq:t-enhancement}
\eea
where in Eq.~(\ref{eq:t-enhancement}) we have neglected the light quark Yukawa couplings.
As expected, in the general case we have three independent flavour phases, in addition to $\theta_L^Y$.
This is representative of the result we  expect in generic NP models, i.e.~models with an extended sector
containing more than one heavy field.
 
Starting from these expressions, we can analyse the specific implementation 
of some of the dynamical alignment mechanisms discussed in Section~\ref{sect:dynamical}
in the context of this model. The condition {\bf II} in Table~\ref{tab:dynamical}, namely 
$\theta_L^{u_1} \approx \theta_L^{u_3}$, is easily obtained in the limit $M^2_\Phi \gg M_{S_1}^2 $. 
As far as the condition  {\bf I} is concerned, we can achieve it neglecting 
the terms with two powers of~$\lambda^R$, and assuming $\lambda^L_{i\alpha}  \ll  \lambda^L_{3\alpha}$ 
for $i=1,2$.
Under reasonable dynamical  assumptions it is therefore easy to reach the condition 
\be
\theta_L^{e\gamma} \approx \theta_L^{u_1} \approx  \theta_L^{u_3} = 
 \frac{{\lambda^L_{31}}^\ast}{{\lambda^L_{32}}^\ast} \,.
 \label{eq:delta_model}
\ee
On the other hand, in this model (as in most models) it is hard to conceive a dynamical alignment of the flavour phase in 
Eq.~(\ref{eq:delta_model}) with the phase of the Yukawa interaction ($\theta_L^Y$).

To further investigate the misalignment of $\theta_L^Y$ and the flavour phase in 
Eq.~(\ref{eq:delta_model}) in this model, it is useful to investigate what happens if we supplement
the dynamical assumptions so far discussed with the hypothesis of a minimally broken $U(2)_{L_L} \times U(2)_{E_R}$ 
symmetry (see Section~\ref{sect:flavsymm}).
To lowest order in the $U(2)_{L_L} \times U(2)_{E_R}$  breaking terms we get 
\be
 \theta_L^{e \gamma} =  \frac{{\lambda^L_{31}}^\ast}{{\lambda^L_{32}}^\ast}  = \frac{ (V_\ell)_1 }{ (V_\ell)_2 }\ \stackrel{ (\ref{eq:standardbasis}) }{\longrightarrow} \ 0 \,,
 \qquad \theta_L^Y = \frac{ (\Delta_e)_{12} }{ (\Delta_e)_{22}  }\ \stackrel{ (\ref{eq:standardbasis}) }{\longrightarrow} \ s_e \,,
\ee
which implies an (unnatural) limit of $\cO(10^{-5})$ on $|s_e|$ in order to satisfy 
the experimental bound in Eq.~(\ref{eq:bound12}). This constraint is much more stringent than the generic condition 
derived in Section~\ref{sect:flavsymm} for minimally broken $U(2)_{L_L} \times U(2)_{E_R}$ models.
This can be understood by noting that after aligning $\theta_L^{e\gamma}$ and $\theta_L^{u_3}$, the leading $U(2)_{L_L} \times U(2)_{E_R}$ spurion contribution to $\theta_L^{e\gamma}$ is suppressed, which is equivalent to assuming a large $b_{e\gamma}/a_{e\gamma}$ ratio in Eq.~(\ref{eq:anbn}).
What was beneficial to reach the dynamical alignment $\theta_L^{e\gamma}  \approx \theta_L^{u_3}$
however has the drawback of making worse the misalignment between  $\theta_L^{e\gamma}$ and~$\theta_L^Y$
compared to the one obtained in a minimally broken $U(2)_{L_L} \times U(2)_{E_R}$ 
framework.


To summarise, in this simplified model it is easy to implement the dynamical hypotheses 
{\bf I} and {\bf II} in Table~\ref{tab:dynamical}. However, by doing so, it becomes more difficult to align the Yukawa 
and dipole interactions. The latter goal is attainable only with the further hypothesis of 
an (almost) exact conservation of the individual lepton numbers.

\section{Conclusion}

The stringent experimental bounds on lepton flavour violating processes involving charged leptons 
indicate that lepton flavour is approximately conserved above the electroweak scale. 
When considering only these bounds, we cannot exclude that this approximate 
symmetry arises accidentally in the SMEFT, being the consequence 
of an overall (flavour anarchic) suppression of $d=6$ operators.  The accidental lepton flavour conservation 
could also arise as the consequence of appropriate scale 
hierarchies, as in the general SM extensions proposed in~\cite{Panico:2016ull,Bordone:2017bld,Allwicher:2020esa,Barbieri:2021wrc}, 
which provide an interesting explanation for the hierarchies observed in both charged-lepton and 
quark Yukawa couplings. 
However, the situation changes if we assume the  $a_\mu$ anomaly 
is due to NP. As we have shown in this paper, in such case we are forced to assume that 
the conservation of lepton flavour, and in particular of the electron flavour, is a key 
property of a large set of operators in the $d=6$ sector of the SMEFT. This property is unlikely to arise accidentally.

The $a_\mu$ anomaly sets a well-defined reference scale (size) for the muon dipole operator, 
with respect to which a series of additional SMEFT operators need to be flavour aligned in order to
satisfy the tight bounds from  $\mu \to e \gamma$ and  $\tau \to \mu \gamma$. 
A consistent treatment of the problem in the SMEFT requires addressing 
the RG evolution not only of the operators mixing into the dipoles, but also of those 
mixing into the effective Yukawa couplings. In general, the five independent flavour phases listed in 
Eq.~(\ref{eq:thetaX}), describing the orientation in flavour space of the corresponding effective operators, 
need to be aligned. In the $e$--$\mu$ case, at least three of these 
phases (dipole, Yukawa, and semileptonic triplet operators) need to be aligned at the $10^{-5}$ level. 
As we have shown in Section~\ref{sect:dynamical}, and illustrated by means of a concrete example 
in Section~\ref{sect:mediators}, dynamical mechanisms can force some of these alignments, but not all of them. 
Barring fine-tuned solutions, the required alignments require extra ingredients, such as 
exact (or almost exact) flavour symmetries ensuring the conservation of the electron flavour.

The most interesting conclusion of this study is that if the  $a_\mu$ anomaly is a sign of NP, 
we are led to conclude that the quark and lepton sectors behave quite differently beyond the SM,
at least in the  few-TeV scale domain, with a lepton sector featuring enhanced symmetries which are not present in the quark sector.

\section*{Acknowledgements}

{We thank Joe Davighi, Admir Greljo, and Anders Eller Thomsen, for useful comments and discussions. 
This project has received funding from the European Research Council (ERC) under the European Union's Horizon 2020 research and innovation programme under grant agreement 833280 (FLAY), and by the Swiss National Science Foundation (SNF) under contract 200020-204428.
 
{\footnotesize
\bibliography{arxiv}

\providecommand{\href}[2]{#2}\begingroup\raggedright\begin{thebibliography}{10}

\bibitem{Muong-2:2021ojo}
{\scshape Muon g-2} collaboration, B.~Abi et~al., \emph{{Measurement of the
  Positive Muon Anomalous Magnetic Moment to 0.46 ppm}},
  \href{http://dx.doi.org/10.1103/PhysRevLett.126.141801}{\emph{Phys. Rev.
  Lett.} {\bf 126} (2021) 141801}, [\href{http://arxiv.org/abs/2104.03281}{{\tt
  2104.03281}}].

\bibitem{Muong-2:2006rrc}
{\scshape Muon g-2} collaboration, G.~W. Bennett et~al., \emph{{Final Report of
  the Muon E821 Anomalous Magnetic Moment Measurement at BNL}},
  \href{http://dx.doi.org/10.1103/PhysRevD.73.072003}{\emph{Phys. Rev. D} {\bf
  73} (2006) 072003}, [\href{http://arxiv.org/abs/hep-ex/0602035}{{\tt
  hep-ex/0602035}}].

\bibitem{Aoyama:2020ynm}
T.~Aoyama et~al., \emph{{The anomalous magnetic moment of the muon in the
  Standard Model}},
  \href{http://dx.doi.org/10.1016/j.physrep.2020.07.006}{\emph{Phys. Rept.}
  {\bf 887} (2020) 1--166}, [\href{http://arxiv.org/abs/2006.04822}{{\tt
  2006.04822}}].

\bibitem{Jegerlehner:2017gek}
F.~Jegerlehner, \emph{{The Anomalous Magnetic Moment of the Muon}}, vol.~274.
\newblock Springer, Cham, 2017.
\newblock 10.1007/978-3-319-63577-4.

\bibitem{Colangelo:2018mtw}
G.~Colangelo, M.~Hoferichter and P.~Stoffer, \emph{{Two-pion contribution to
  hadronic vacuum polarization}},
  \href{http://dx.doi.org/10.1007/JHEP02(2019)006}{\emph{JHEP} {\bf 02} (2019)
  006}, [\href{http://arxiv.org/abs/1810.00007}{{\tt 1810.00007}}].

\bibitem{Hoferichter:2019mqg}
M.~Hoferichter, B.-L. Hoid and B.~Kubis, \emph{{Three-pion contribution to
  hadronic vacuum polarization}},
  \href{http://dx.doi.org/10.1007/JHEP08(2019)137}{\emph{JHEP} {\bf 08} (2019)
  137}, [\href{http://arxiv.org/abs/1907.01556}{{\tt 1907.01556}}].

\bibitem{Davier:2019can}
M.~Davier, A.~Hoecker, B.~Malaescu and Z.~Zhang, \emph{{A new evaluation of the
  hadronic vacuum polarisation contributions to the muon anomalous magnetic
  moment and to ${\alpha(m_Z^2)}$}},
  \href{http://dx.doi.org/10.1140/epjc/s10052-020-7792-2}{\emph{Eur. Phys. J.
  C} {\bf 80} (2020) 241}, [\href{http://arxiv.org/abs/1908.00921}{{\tt
  1908.00921}}].

\bibitem{Keshavarzi:2019abf}
A.~Keshavarzi, D.~Nomura and T.~Teubner, \emph{{$g-2$ of charged leptons,
  $\alpha (M^2_Z)$ , and the hyperfine splitting of muonium}},
  \href{http://dx.doi.org/10.1103/PhysRevD.101.014029}{\emph{Phys. Rev. D} {\bf
  101} (2020) 014029}, [\href{http://arxiv.org/abs/1911.00367}{{\tt
  1911.00367}}].

\bibitem{Hoid:2020xjs}
B.-L. Hoid, M.~Hoferichter and B.~Kubis, \emph{{Hadronic vacuum polarization
  and vector-meson resonance parameters from $e^+e^-\rightarrow \pi
  ^0\gamma$}},
  \href{http://dx.doi.org/10.1140/epjc/s10052-020-08550-2}{\emph{Eur. Phys. J.
  C} {\bf 80} (2020) 988}, [\href{http://arxiv.org/abs/2007.12696}{{\tt
  2007.12696}}].

\bibitem{Czarnecki:2002nt}
A.~Czarnecki, W.~J. Marciano and A.~Vainshtein, \emph{{Refinements in
  electroweak contributions to the muon anomalous magnetic moment}},
  \href{http://dx.doi.org/10.1103/PhysRevD.67.073006}{\emph{Phys. Rev. D} {\bf
  67} (2003) 073006}, [\href{http://arxiv.org/abs/hep-ph/0212229}{{\tt
  hep-ph/0212229}}].

\bibitem{Melnikov:2003xd}
K.~Melnikov and A.~Vainshtein, \emph{{Hadronic light-by-light scattering
  contribution to the muon anomalous magnetic moment revisited}},
  \href{http://dx.doi.org/10.1103/PhysRevD.70.113006}{\emph{Phys. Rev. D} {\bf
  70} (2004) 113006}, [\href{http://arxiv.org/abs/hep-ph/0312226}{{\tt
  hep-ph/0312226}}].

\bibitem{Aoyama:2012wk}
T.~Aoyama, M.~Hayakawa, T.~Kinoshita and M.~Nio, \emph{{Complete Tenth-Order
  QED Contribution to the Muon g-2}},
  \href{http://dx.doi.org/10.1103/PhysRevLett.109.111808}{\emph{Phys. Rev.
  Lett.} {\bf 109} (2012) 111808}, [\href{http://arxiv.org/abs/1205.5370}{{\tt
  1205.5370}}].

\bibitem{Gnendiger:2013pva}
C.~Gnendiger, D.~St\"ockinger and H.~St\"ockinger-Kim, \emph{{The electroweak
  contributions to $(g-2)_\mu$ after the Higgs boson mass measurement}},
  \href{http://dx.doi.org/10.1103/PhysRevD.88.053005}{\emph{Phys. Rev. D} {\bf
  88} (2013) 053005}, [\href{http://arxiv.org/abs/1306.5546}{{\tt 1306.5546}}].

\bibitem{Borsanyi:2020mff}
S.~Borsanyi et~al., \emph{{Leading hadronic contribution to the muon magnetic
  moment from lattice QCD}},
  \href{http://dx.doi.org/10.1038/s41586-021-03418-1}{\emph{Nature} {\bf 593}
  (2021) 51--55}, [\href{http://arxiv.org/abs/2002.12347}{{\tt 2002.12347}}].

\bibitem{Aebischer:2021uvt}
J.~Aebischer, W.~Dekens, E.~E. Jenkins, A.~V. Manohar, D.~Sengupta and
  P.~Stoffer, \emph{{Effective field theory interpretation of lepton magnetic
  and electric dipole moments}},
  \href{http://dx.doi.org/10.1007/JHEP07(2021)107}{\emph{JHEP} {\bf 07} (2021)
  107}, [\href{http://arxiv.org/abs/2102.08954}{{\tt 2102.08954}}].

\bibitem{Allwicher_2021}
L.~Allwicher, L.~D. Luzio, M.~Fedele, F.~Mescia and M.~Nardecchia, \emph{What
  is the scale of new physics behind the muon g-2 ?},
  \href{http://dx.doi.org/10.1103/physrevd.104.055035}{\emph{Physical Review D}
  {\bf 104} (sep, 2021) }.

\bibitem{Allwicher:2021rtd}
L.~Allwicher, P.~Arnan, D.~Barducci and M.~Nardecchia, \emph{{Perturbative
  unitarity constraints on generic Yukawa interactions}},
  \href{http://dx.doi.org/10.1007/JHEP10(2021)129}{\emph{JHEP} {\bf 10} (2021)
  129}, [\href{http://arxiv.org/abs/2108.00013}{{\tt 2108.00013}}].

\bibitem{Fajfer:2021cxa}
S.~Fajfer, J.~F. Kamenik and M.~Tammaro, \emph{{Interplay of New Physics
  effects in $(g-2)_\ell$ and $h \to \ell^+ \ell^-$ lessons from SMEFT}},
  \href{http://dx.doi.org/10.1007/JHEP06(2021)099}{\emph{JHEP} {\bf 06} (2021)
  099}, [\href{http://arxiv.org/abs/2103.10859}{{\tt 2103.10859}}].

\bibitem{LHCb:2014vgu}
{\scshape LHCb} collaboration, R.~Aaij et~al., \emph{{Test of lepton
  universality using $B^{+}\rightarrow K^{+}\ell^{+}\ell^{-}$ decays}},
  \href{http://dx.doi.org/10.1103/PhysRevLett.113.151601}{\emph{Phys. Rev.
  Lett.} {\bf 113} (2014) 151601}, [\href{http://arxiv.org/abs/1406.6482}{{\tt
  1406.6482}}].

\bibitem{LHCb:2017avl}
{\scshape LHCb} collaboration, R.~Aaij et~al., \emph{{Test of lepton
  universality with $B^{0} \rightarrow K^{*0}\ell^{+}\ell^{-}$ decays}},
  \href{http://dx.doi.org/10.1007/JHEP08(2017)055}{\emph{JHEP} {\bf 08} (2017)
  055}, [\href{http://arxiv.org/abs/1705.05802}{{\tt 1705.05802}}].

\bibitem{LHCb:2019hip}
{\scshape LHCb} collaboration, R.~Aaij et~al., \emph{{Search for
  lepton-universality violation in $B^+\to K^+\ell^+\ell^-$ decays}},
  \href{http://dx.doi.org/10.1103/PhysRevLett.122.191801}{\emph{Phys. Rev.
  Lett.} {\bf 122} (2019) 191801}, [\href{http://arxiv.org/abs/1903.09252}{{\tt
  1903.09252}}].

\bibitem{LHCb:2021trn}
{\scshape LHCb} collaboration, R.~Aaij et~al., \emph{{Test of lepton
  universality in beauty-quark decays}},
  \href{http://arxiv.org/abs/2103.11769}{{\tt 2103.11769}}.

\bibitem{BaBar:2012obs}
{\scshape BaBar} collaboration, J.~P. Lees et~al., \emph{{Evidence for an
  excess of $\bar{B} \to D^{(*)} \tau^-\bar{\nu}_\tau$ decays}},
  \href{http://dx.doi.org/10.1103/PhysRevLett.109.101802}{\emph{Phys. Rev.
  Lett.} {\bf 109} (2012) 101802}, [\href{http://arxiv.org/abs/1205.5442}{{\tt
  1205.5442}}].

\bibitem{BaBar:2013mob}
{\scshape BaBar} collaboration, J.~P. Lees et~al., \emph{{Measurement of an
  Excess of $\bar{B} \to D^{(*)}\tau^- \bar{\nu}_\tau$ Decays and Implications
  for Charged Higgs Bosons}},
  \href{http://dx.doi.org/10.1103/PhysRevD.88.072012}{\emph{Phys. Rev. D} {\bf
  88} (2013) 072012}, [\href{http://arxiv.org/abs/1303.0571}{{\tt 1303.0571}}].

\bibitem{Belle:2015qfa}
{\scshape Belle} collaboration, M.~Huschle et~al., \emph{{Measurement of the
  branching ratio of $\bar{B} \to D^{(\ast)} \tau^- \bar{\nu}_\tau$ relative to
  $\bar{B} \to D^{(\ast)} \ell^- \bar{\nu}_\ell$ decays with hadronic tagging
  at Belle}}, \href{http://dx.doi.org/10.1103/PhysRevD.92.072014}{\emph{Phys.
  Rev. D} {\bf 92} (2015) 072014}, [\href{http://arxiv.org/abs/1507.03233}{{\tt
  1507.03233}}].

\bibitem{LHCb:2015gmp}
{\scshape LHCb} collaboration, R.~Aaij et~al., \emph{{Measurement of the ratio
  of branching fractions $\mathcal{B}(\bar{B}^0 \to
  D^{*+}\tau^{-}\bar{\nu}_{\tau})/\mathcal{B}(\bar{B}^0 \to
  D^{*+}\mu^{-}\bar{\nu}_{\mu})$}},
  \href{http://dx.doi.org/10.1103/PhysRevLett.115.111803}{\emph{Phys. Rev.
  Lett.} {\bf 115} (2015) 111803}, [\href{http://arxiv.org/abs/1506.08614}{{\tt
  1506.08614}}].

\bibitem{LHCb:2017smo}
{\scshape LHCb} collaboration, R.~Aaij et~al., \emph{{Measurement of the ratio
  of the $B^0 \to D^{*-} \tau^+ \nu_{\tau}$ and $B^0 \to D^{*-} \mu^+
  \nu_{\mu}$ branching fractions using 3-prong $\tau$-lepton decays}},
  \href{http://dx.doi.org/10.1103/PhysRevLett.120.171802}{\emph{Phys. Rev.
  Lett.} {\bf 120} (2018) 171802}, [\href{http://arxiv.org/abs/1708.08856}{{\tt
  1708.08856}}].

\bibitem{LHCb:2017rln}
{\scshape LHCb} collaboration, R.~Aaij et~al., \emph{{Test of Lepton Flavor
  Universality by the measurement of the $B^0 \to D^{*-} \tau^+ \nu_{\tau}$
  branching fraction using 3-prong $\tau$ decays}},
  \href{http://dx.doi.org/10.1103/PhysRevD.97.072013}{\emph{Phys. Rev. D} {\bf
  97} (2018) 072013}, [\href{http://arxiv.org/abs/1711.02505}{{\tt
  1711.02505}}].

\bibitem{Greljo:2021xmg}
A.~Greljo, P.~Stangl and A.~E. Thomsen, \emph{{A model of muon anomalies}},
  \href{http://dx.doi.org/10.1016/j.physletb.2021.136554}{\emph{Phys. Lett. B}
  {\bf 820} (2021) 136554}, [\href{http://arxiv.org/abs/2103.13991}{{\tt
  2103.13991}}].

\bibitem{Baum:2021qzx}
S.~Baum, M.~Carena, N.~R. Shah and C.~E.~M. Wagner, \emph{{The Tiny (g-2) Muon
  Wobble from Small-$\mu$ Supersymmetry}},
  \href{http://arxiv.org/abs/2104.03302}{{\tt 2104.03302}}.

\bibitem{Lee:2021jdr}
H.~M. Lee, \emph{{Leptoquark option for B-meson anomalies and leptonic
  signatures}},
  \href{http://dx.doi.org/10.1103/PhysRevD.104.015007}{\emph{Phys. Rev. D} {\bf
  104} (2021) 015007}, [\href{http://arxiv.org/abs/2104.02982}{{\tt
  2104.02982}}].

\bibitem{Arcadi:2021cwg}
G.~Arcadi, L.~Calibbi, M.~Fedele and F.~Mescia, \emph{{Muon $g-2$ and
  $B$-anomalies from Dark Matter}},
  \href{http://dx.doi.org/10.1103/PhysRevLett.127.061802}{\emph{Phys. Rev.
  Lett.} {\bf 127} (2021) 061802}, [\href{http://arxiv.org/abs/2104.03228}{{\tt
  2104.03228}}].

\bibitem{Cen:2021iwv}
J.-Y. Cen, Y.~Cheng, X.-G. He and J.~Sun, \emph{{Flavor Specific
  $U(1)_{B_q-L_\mu}$ Gauge Model for Muon $g-2$ and $b \to s \bar \mu \mu$
  Anomalies}},  \href{http://arxiv.org/abs/2104.05006}{{\tt 2104.05006}}.

\bibitem{Marzocca:2021azj}
D.~Marzocca and S.~Trifinopoulos, \emph{{Minimal Explanation of Flavor
  Anomalies: B-Meson Decays, Muon Magnetic Moment, and the Cabibbo Angle}},
  \href{http://dx.doi.org/10.1103/PhysRevLett.127.061803}{\emph{Phys. Rev.
  Lett.} {\bf 127} (2021) 061803}, [\href{http://arxiv.org/abs/2104.05730}{{\tt
  2104.05730}}].

\bibitem{Altmannshofer:2021hfu}
W.~Altmannshofer, S.~A. Gadam, S.~Gori and N.~Hamer, \emph{{Explaining
  $(g-2)_{\mu}$ with Multi-TeV Sleptons}},
  \href{http://arxiv.org/abs/2104.08293}{{\tt 2104.08293}}.

\bibitem{Cacciapaglia:2021gff}
G.~Cacciapaglia, C.~Cot and F.~Sannino, \emph{{Naturalness of lepton
  non-universality and muon g-2}},  \href{http://arxiv.org/abs/2104.08818}{{\tt
  2104.08818}}.

\bibitem{Davighi:2021oel}
J.~Davighi, \emph{{Anomalous Z' bosons for anomalous B decays}},
  \href{http://dx.doi.org/10.1007/JHEP08(2021)101}{\emph{JHEP} {\bf 08} (2021)
  101}, [\href{http://arxiv.org/abs/2105.06918}{{\tt 2105.06918}}].

\bibitem{Marzocca:2021miv}
D.~Marzocca, S.~Trifinopoulos and E.~Venturini, \emph{{From B-meson anomalies
  to Kaon physics with scalar leptoquarks}},
  \href{http://arxiv.org/abs/2106.15630}{{\tt 2106.15630}}.

\bibitem{Greljo:2021npi}
A.~Greljo, Y.~Soreq, P.~Stangl, A.~E. Thomsen and J.~Zupan, \emph{{Muonic Force
  Behind Flavor Anomalies}},  \href{http://arxiv.org/abs/2107.07518}{{\tt
  2107.07518}}.

\bibitem{Bause:2021prv}
R.~Bause, G.~Hiller, T.~H\"ohne, D.~F. Litim and T.~Steudtner,
  \emph{{B-Anomalies from flavorful U(1)' extensions, safely}},
  \href{http://arxiv.org/abs/2109.06201}{{\tt 2109.06201}}.

\bibitem{Buttazzo:2020eyl}
D.~Buttazzo and P.~Paradisi, \emph{{Probing the muon g-2 anomaly at a Muon
  Collider}},  \href{http://arxiv.org/abs/2012.02769}{{\tt 2012.02769}}.

\bibitem{MEG:2016leq}
{\scshape MEG} collaboration, A.~M. Baldini et~al., \emph{{Search for the
  lepton flavour violating decay $\mu ^+ \rightarrow \mathrm {e}^+ \gamma $
  with the full dataset of the MEG experiment}},
  \href{http://dx.doi.org/10.1140/epjc/s10052-016-4271-x}{\emph{Eur. Phys. J.
  C} {\bf 76} (2016) 434}, [\href{http://arxiv.org/abs/1605.05081}{{\tt
  1605.05081}}].

\bibitem{BaBar:2009hkt}
{\scshape BaBar} collaboration, B.~Aubert et~al., \emph{{Searches for Lepton
  Flavor Violation in the Decays $\tau^\pm \to e^\pm \gamma$ and $\tau^\pm \to
  \mu^\pm \gamma$}},
  \href{http://dx.doi.org/10.1103/PhysRevLett.104.021802}{\emph{Phys. Rev.
  Lett.} {\bf 104} (2010) 021802}, [\href{http://arxiv.org/abs/0908.2381}{{\tt
  0908.2381}}].

\bibitem{Grzadkowski:2010es}
B.~Grzadkowski, M.~Iskrzynski, M.~Misiak and J.~Rosiek, \emph{{Dimension-Six
  Terms in the Standard Model Lagrangian}},
  \href{http://dx.doi.org/10.1007/JHEP10(2010)085}{\emph{JHEP} {\bf 10} (2010)
  085}, [\href{http://arxiv.org/abs/1008.4884}{{\tt 1008.4884}}].

\bibitem{Jenkins:2013zja}
E.~E. Jenkins, A.~V. Manohar and M.~Trott, \emph{{Renormalization Group
  Evolution of the Standard Model Dimension Six Operators I: Formalism and
  {$\lambda$} Dependence}},
  \href{http://dx.doi.org/10.1007/JHEP10(2013)087}{\emph{JHEP} {\bf 10} (2013)
  087}, [\href{http://arxiv.org/abs/1308.2627}{{\tt 1308.2627}}].

\bibitem{Jenkins:2013wua}
E.~E. Jenkins, A.~V. Manohar and M.~Trott, \emph{{Renormalization Group
  Evolution of the Standard Model Dimension Six Operators II: Yukawa
  Dependence}}, \href{http://dx.doi.org/10.1007/JHEP01(2014)035}{\emph{JHEP}
  {\bf 01} (2014) 035}, [\href{http://arxiv.org/abs/1310.4838}{{\tt
  1310.4838}}].

\bibitem{Alonso:2013hga}
R.~Alonso, E.~E. Jenkins, A.~V. Manohar and M.~Trott, \emph{{RG Evolution of
  the Standard Model Dimension Six Operators III: Gauge Coupling Dependence and
  Phenomenology}}, \href{http://dx.doi.org/10.1007/JHEP04(2014)159}{\emph{JHEP}
  {\bf 04} (2014) 159}, [\href{http://arxiv.org/abs/1312.2014}{{\tt
  1312.2014}}].

\bibitem{Celis:2017hod}
A.~Celis, J.~Fuentes-Martin, A.~Vicente and J.~Virto, \emph{{DsixTools: The
  Standard Model Effective Field Theory Toolkit}},
  \href{http://dx.doi.org/10.1140/epjc/s10052-017-4967-6}{\emph{Eur. Phys. J.
  C} {\bf 77} (2017) 405}, [\href{http://arxiv.org/abs/1704.04504}{{\tt
  1704.04504}}].

\bibitem{Blankenburg:2012ex}
G.~Blankenburg, J.~Ellis and G.~Isidori, \emph{{Flavour-Changing Decays of a
  125 GeV Higgs-like Particle}},
  \href{http://dx.doi.org/10.1016/j.physletb.2012.05.007}{\emph{Phys. Lett. B}
  {\bf 712} (2012) 386--390}, [\href{http://arxiv.org/abs/1202.5704}{{\tt
  1202.5704}}].

\bibitem{Harnik:2012pb}
R.~Harnik, J.~Kopp and J.~Zupan, \emph{{Flavor Violating Higgs Decays}},
  \href{http://dx.doi.org/10.1007/JHEP03(2013)026}{\emph{JHEP} {\bf 03} (2013)
  026}, [\href{http://arxiv.org/abs/1209.1397}{{\tt 1209.1397}}].

\bibitem{Marciano:2016yhf}
W.~J. Marciano, A.~Masiero, P.~Paradisi and M.~Passera, \emph{{Contributions of
  axionlike particles to lepton dipole moments}},
  \href{http://dx.doi.org/10.1103/PhysRevD.94.115033}{\emph{Phys. Rev. D} {\bf
  94} (2016) 115033}, [\href{http://arxiv.org/abs/1607.01022}{{\tt
  1607.01022}}].

\bibitem{Bauer:2019gfk}
M.~Bauer, M.~Neubert, S.~Renner, M.~Schnubel and A.~Thamm, \emph{{Axionlike
  Particles, Lepton-Flavor Violation, and a New Explanation of $a_\mu$ and
  $a_e$}}, \href{http://dx.doi.org/10.1103/PhysRevLett.124.211803}{\emph{Phys.
  Rev. Lett.} {\bf 124} (2020) 211803},
  [\href{http://arxiv.org/abs/1908.00008}{{\tt 1908.00008}}].

\bibitem{Cornella:2019uxs}
C.~Cornella, P.~Paradisi and O.~Sumensari, \emph{{Hunting for ALPs with Lepton
  Flavor Violation}},
  \href{http://dx.doi.org/10.1007/JHEP01(2020)158}{\emph{JHEP} {\bf 01} (2020)
  158}, [\href{http://arxiv.org/abs/1911.06279}{{\tt 1911.06279}}].

\bibitem{Buen-Abad:2021fwq}
M.~A. Buen-Abad, J.~Fan, M.~Reece and C.~Sun, \emph{{Challenges for an axion
  explanation of the muon $g - 2$ measurement}},
  \href{http://dx.doi.org/10.1007/JHEP09(2021)101}{\emph{JHEP} {\bf 09} (2021)
  101}, [\href{http://arxiv.org/abs/2104.03267}{{\tt 2104.03267}}].

\bibitem{Brdar:2021pla}
V.~Brdar, S.~Jana, J.~Kubo and M.~Lindner, \emph{{Semi-secretly interacting
  Axion-like particle as an explanation of Fermilab muon g?\ensuremath{-}?2
  measurement}},
  \href{http://dx.doi.org/10.1016/j.physletb.2021.136529}{\emph{Phys. Lett. B}
  {\bf 820} (2021) 136529}, [\href{http://arxiv.org/abs/2104.03282}{{\tt
  2104.03282}}].

\bibitem{Baker:2021yli}
M.~J. Baker, P.~Cox and R.~R. Volkas, \emph{{Radiative muon mass models and
  $(g-2)_\mu$}}, \href{http://dx.doi.org/10.1007/JHEP05(2021)174}{\emph{JHEP}
  {\bf 05} (2021) 174}, [\href{http://arxiv.org/abs/2103.13401}{{\tt
  2103.13401}}].

\bibitem{Yin:2021yqy}
W.~Yin, \emph{{Radiative lepton mass and muon $g-2$ with suppressed lepton
  flavor and CP violations}},  \href{http://arxiv.org/abs/2103.14234}{{\tt
  2103.14234}}.

\bibitem{Gabrielli:2013jka}
E.~Gabrielli and M.~Raidal, \emph{{Exponentially spread dynamical Yukawa
  couplings from nonperturbative chiral symmetry breaking in the dark sector}},
  \href{http://dx.doi.org/10.1103/PhysRevD.89.015008}{\emph{Phys. Rev. D} {\bf
  89} (2014) 015008}, [\href{http://arxiv.org/abs/1310.1090}{{\tt 1310.1090}}].

\bibitem{Gabrielli:2016cut}
E.~Gabrielli, B.~Mele, M.~Raidal and E.~Venturini, \emph{{FCNC decays of
  standard model fermions into a dark photon}},
  \href{http://dx.doi.org/10.1103/PhysRevD.94.115013}{\emph{Phys. Rev. D} {\bf
  94} (2016) 115013}, [\href{http://arxiv.org/abs/1607.05928}{{\tt
  1607.05928}}].

\bibitem{Barbieri:2011ci}
R.~Barbieri, G.~Isidori, J.~Jones-Perez, P.~Lodone and D.~M. Straub,
  \emph{{$U(2)$ and Minimal Flavour Violation in Supersymmetry}},
  \href{http://dx.doi.org/10.1140/epjc/s10052-011-1725-z}{\emph{Eur. Phys. J.
  C} {\bf 71} (2011) 1725}, [\href{http://arxiv.org/abs/1105.2296}{{\tt
  1105.2296}}].

\bibitem{Barbieri:2012uh}
R.~Barbieri, D.~Buttazzo, F.~Sala and D.~M. Straub, \emph{{Flavour physics from
  an approximate $U(2)^3$ symmetry}},
  \href{http://dx.doi.org/10.1007/JHEP07(2012)181}{\emph{JHEP} {\bf 07} (2012)
  181}, [\href{http://arxiv.org/abs/1203.4218}{{\tt 1203.4218}}].

\bibitem{Blankenburg:2012nx}
G.~Blankenburg, G.~Isidori and J.~Jones-Perez, \emph{{Neutrino Masses and LFV
  from Minimal Breaking of $U(3)^5$ and $U(2)^5$ flavor Symmetries}},
  \href{http://dx.doi.org/10.1140/epjc/s10052-012-2126-7}{\emph{Eur. Phys. J.
  C} {\bf 72} (2012) 2126}, [\href{http://arxiv.org/abs/1204.0688}{{\tt
  1204.0688}}].

\bibitem{Greljo:2015mma}
A.~Greljo, G.~Isidori and D.~Marzocca, \emph{{On the breaking of Lepton Flavor
  Universality in B decays}},
  \href{http://dx.doi.org/10.1007/JHEP07(2015)142}{\emph{JHEP} {\bf 07} (2015)
  142}, [\href{http://arxiv.org/abs/1506.01705}{{\tt 1506.01705}}].

\bibitem{Barbieri:2015yvd}
R.~Barbieri, G.~Isidori, A.~Pattori and F.~Senia, \emph{{Anomalies in
  $B$-decays and $U(2)$ flavour symmetry}},
  \href{http://dx.doi.org/10.1140/epjc/s10052-016-3905-3}{\emph{Eur. Phys. J.
  C} {\bf 76} (2016) 67}, [\href{http://arxiv.org/abs/1512.01560}{{\tt
  1512.01560}}].

\bibitem{Bordone:2017anc}
M.~Bordone, G.~Isidori and S.~Trifinopoulos, \emph{{Semileptonic $B$-physics
  anomalies: A general EFT analysis within $U(2)^n$ flavor symmetry}},
  \href{http://dx.doi.org/10.1103/PhysRevD.96.015038}{\emph{Phys. Rev. D} {\bf
  96} (2017) 015038}, [\href{http://arxiv.org/abs/1702.07238}{{\tt
  1702.07238}}].

\bibitem{Fuentes-Martin:2019mun}
J.~Fuentes-Mart\'\i{}n, G.~Isidori, J.~Pag\`es and K.~Yamamoto, \emph{{With or
  without U(2)? Probing non-standard flavor and helicity structures in
  semileptonic B decays}},
  \href{http://dx.doi.org/10.1016/j.physletb.2019.135080}{\emph{Phys. Lett. B}
  {\bf 800} (2020) 135080}, [\href{http://arxiv.org/abs/1909.02519}{{\tt
  1909.02519}}].

\bibitem{Cornella:2021sby}
C.~Cornella, D.~A. Faroughy, J.~Fuentes-Martin, G.~Isidori and M.~Neubert,
  \emph{{Reading the footprints of the B-meson flavor anomalies}},
  \href{http://dx.doi.org/10.1007/JHEP08(2021)050}{\emph{JHEP} {\bf 08} (2021)
  050}, [\href{http://arxiv.org/abs/2103.16558}{{\tt 2103.16558}}].

\bibitem{Faroughy:2020ina}
D.~A. Faroughy, G.~Isidori, F.~Wilsch and K.~Yamamoto, \emph{{Flavour
  symmetries in the SMEFT}},
  \href{http://dx.doi.org/10.1007/JHEP08(2020)166}{\emph{JHEP} {\bf 08} (2020)
  166}, [\href{http://arxiv.org/abs/2005.05366}{{\tt 2005.05366}}].

\bibitem{Gherardi:2020det}
V.~Gherardi, D.~Marzocca and E.~Venturini, \emph{{Matching scalar leptoquarks
  to the SMEFT at one loop}},
  \href{http://dx.doi.org/10.1007/JHEP07(2020)225}{\emph{JHEP} {\bf 07} (2020)
  225}, [\href{http://arxiv.org/abs/2003.12525}{{\tt 2003.12525}}].

\bibitem{Dorsner:2020aaz}
I.~Dor\v{s}ner, S.~Fajfer and S.~Saad, \emph{{$\mu \to e \gamma$ selecting
  scalar leptoquark solutions for the $(g-2)_{e,\mu}$ puzzles}},
  \href{http://dx.doi.org/10.1103/PhysRevD.102.075007}{\emph{Phys. Rev. D} {\bf
  102} (2020) 075007}, [\href{http://arxiv.org/abs/2006.11624}{{\tt
  2006.11624}}].

\bibitem{Dedes:2021abc}
A.~Dedes and K.~Mantzaropoulos, \emph{{Universal Scalar Leptoquark Action for
  Matching}},  \href{http://arxiv.org/abs/2108.10055}{{\tt 2108.10055}}.

\bibitem{Panico:2016ull}
G.~Panico and A.~Pomarol, \emph{{Flavor hierarchies from dynamical scales}},
  \href{http://dx.doi.org/10.1007/JHEP07(2016)097}{\emph{JHEP} {\bf 07} (2016)
  097}, [\href{http://arxiv.org/abs/1603.06609}{{\tt 1603.06609}}].

\bibitem{Bordone:2017bld}
M.~Bordone, C.~Cornella, J.~Fuentes-Martin and G.~Isidori, \emph{{A three-site
  gauge model for flavor hierarchies and flavor anomalies}},
  \href{http://dx.doi.org/10.1016/j.physletb.2018.02.011}{\emph{Phys. Lett. B}
  {\bf 779} (2018) 317--323}, [\href{http://arxiv.org/abs/1712.01368}{{\tt
  1712.01368}}].

\bibitem{Allwicher:2020esa}
L.~Allwicher, G.~Isidori and A.~E. Thomsen, \emph{{Stability of the Higgs
  Sector in a Flavor-Inspired Multi-Scale Model}},
  \href{http://dx.doi.org/10.1007/JHEP01(2021)191}{\emph{JHEP} {\bf 01} (2021)
  191}, [\href{http://arxiv.org/abs/2011.01946}{{\tt 2011.01946}}].

\bibitem{Barbieri:2021wrc}
R.~Barbieri, \emph{{A View of Flavour Physics in 2021}},
  \href{http://dx.doi.org/10.5506/APhysPolB.52.789}{\emph{Acta Phys. Polon. B}
  {\bf 52} (2021) 789}, [\href{http://arxiv.org/abs/2103.15635}{{\tt
  2103.15635}}].

\end{thebibliography}\endgroup
}


\end{document}